\documentclass[namedreferences,hyperref,linksfromyear,optionalrh,solaromanenum]{spr-sola}

\hypersetup{colorlinks=true, linkcolor=blue, citecolor=blue}
\usepackage{graphicx}
\usepackage{natbib}
\usepackage{multirow}
\usepackage{txfonts}
\usepackage{siunitx}

\begin{document}

\begin{frontmatter}

\title{Polarity Reversal of the Polar Magnetic Fields in Solar Cycle 25}

\author[addressref={1,2},email={}]{\inits{Y. }\fnm{Yin } \snm{Li}\orcid{0009-0003-2600-4835}}
\author[addressref={1,2},corref,email={shuhongyang@nao.cas.cn}]{\inits{S.H. }\fnm{Shuhong }\snm{Yang}\orcid{0000-0002-6565-3251}}
\author[addressref={1,2},email={}]{\inits{Y.Z. }\fnm{Yuzong }\snm{Zhang}\orcid{0000-0002-2869-0651}}
\author[addressref={3,4,5},email={}]{\inits{Q. }\fnm{Qiao }\snm{Song}\orcid{0000-0003-3568-445X}}
\author[addressref={1,2},email={}]{\inits{G.P. }\fnm{Guiping }\snm{Zhou}\orcid{0000-0001-8228-565X}}
\author[addressref={1,2},email={}]{\inits{Y.Y. }\fnm{Yuanyong }\snm{Deng}\orcid{0000-0003-1988-4574}}
\author[addressref={1,2},email={}]{\inits{J.X. }\fnm{Jingxiu }\snm{Wang}\orcid{0000-0003-2544-9544}}

\runningauthor{}
\runningtitle{}

\address[id={1}]{State Key Laboratory of Solar Activity and Space Weather, National Astronomical Observatories, Chinese Academy of Sciences, Beijing 100101, People's Republic of China}
\address[id={2}]{School of Astronomy and Space Science, University of Chinese Academy of Sciences, Beijing 100049, People's Republic of China}
\address[id={3}]{National Satellite Meteorological Center (National Centre for Space Weather), China Meteorological Administration, Beijing 100081, People's Republic of China}
\address[id={4}]{Innovation Center for FengYun Meteorological Satellite (FYSIC), Beijing 100081, People's Republic of China}
\address[id={5}]{Key Laboratory of Space Weather, China Meteorological Administration, Beijing 100081, People's Republic of China}

\begin{abstract}
The polar magnetic field polarity reversal is a key signature of solar cycle evolution, and precise determination of its timing is crucial for dynamo theory validation and solar cycle prediction. We investigate the polar polarity reversal of solar cycle 25 using the vector magnetic field data from the spectropolarimeter on board the Hinode satellite. We constructed polar top-down composite maps from Hinode-view magnetograms. These maps show the year-to-year polar polarity variations, with the northern polar region gradually changing from positive to negative and the southern polar region exhibiting the reverse behavior. The polarity reversals of the northern and southern polar caps (above $70^\circ$ latitude) likely occurred in November 2024 and October 2024, respectively. The northern polarity reversal lagged the northern hemispheric sunspot number maximum by approximately 19 months, while the southern reversal possibly coincided with the southern maximum. Moreover, polarity reversal times calculated at $5^\circ$ latitude intervals above $70^\circ$ reveal a trend of earlier reversal in lower latitudes consistent with that of solar cycle 24. These results offer observational references for modeling polar polarity reversal in solar cycles.
\end{abstract}
\end{frontmatter}

\section{Introduction}
The polarity reversal of the solar polar magnetic fields plays a critical role in deciphering the origin of solar magnetism and cyclic variations \citep{mordvinov_reversals_2014,petrie_solar_2015}. \citet{babcock_suns_1959} was the first to discover the polarity reversal of the magnetic field near the pole. Some theoretical models, e.g. surface flux transport (SFT; \citealt{leighton_transport_1964,jiang_magnetic_2014}) model, replicate that polar magnetic field polarity reversal occurs around the sunspot maximum phase of each 11-year solar cycle (SC), driven by the combined effects of differential rotation, meridional circulation, granulation and supergranulation \citep{leighton_transport_1964}. Polar magnetic fields at the minimum phase of SC provides the seed field for the next cycle \citep{petrovay_solar_2020}. \cite{charbonneau_dynamo_2020} pointed out that a model of solar dynamo should reproduce the SC polarity reversals and polar field strength. Conversely, the properties of polar magnetic field could exert constraints on dynamo model \citep{jiang_magnetic_2014} and be taken as precursors for predicting the amplitude of the next cycle \citep{cameron_solar_2016, kumar_polar_2021}.

\begin{table}
\caption{Datasets of Hinode/SP used in this study}
\label{tabel1}
\begin{tabular}{ccccccc}
\hline
\multicolumn{3}{c}{South} & & \multicolumn{3}{c}{North} \\
\hline
\multirow{2}{*}{Date} & \multicolumn{2}{c}{Time (UT)} & & \multirow{2}{*}{Date} & \multicolumn{2}{c}{Time (UT)} \\
\cline{2-3} \cline{6-7}
& Start & End & & & Start & End \\
\hline
2022-03-07 & 11:14:04 & 14:07:34 & & 2022-08-22 & 12:02:05 & 14:55:33 \\
2022-03-10 & 15:00:06 & 17:53:35 & & 2022-08-25 & 14:30:05 & 17:23:34 \\
2022-03-13 & 12:04:05 & 14:56:23 & & 2022-08-28 & 13:02:46 & 15:56:15 \\
2022-03-16 & 13:49:05 & 16:42:34 & & 2022-08-31 & 12:05:52 & 14:59:22 \\
2022-03-19 & 15:00:05 & 17:53:34 & & 2022-09-03 & 12:11:45 & 15:05:14 \\
2022-03-22 & 14:45:06 & 17:38:34 & & 2022-09-06 & 11:17:51 & 14:11:19 \\
2022-04-07 & 11:36:05 & 14:29:34 & & 2022-09-09 & 12:05:46 & 14:59:15 \\
 & & & & 2022-09-12 & 11:04:32 & 13:58:00 \\
 & & & & 2022-09-15 & 14:40:05 & 17:33:34 \\
\hline
2023-02-23 & 11:45:56 & 14:39:24 & & 2023-08-23 & 14:10:04 & 17:03:33 \\
2023-02-26 & 11:59:55 & 14:53:23 & & 2023-08-29 & 11:04:03 & 13:57:33 \\
2023-03-01 & 10:34:56 & 13:28:25 & & 2023-09-01 & 13:30:05 & 16:23:34 \\
2023-03-04 & 11:02:57 & 13:56:26 & & 2023-09-04 & 14:10:04 & 17:03:32 \\
2023-03-07 & 11:00:33 & 13:54:01 & & 2023-09-10 & 12:35:31 & 15:28:59 \\
2023-03-10 & 10:56:13 & 13:49:41 & & 2023-09-13 & 11:21:32 & 14:15:00 \\
2023-03-13 & 10:04:31 & 12:58:00 & & 2023-09-19 & 14:30:05 & 17:23:35 \\
2023-03-16 & 15:13:04 & 18:06:33 & & 2023-09-22 & 14:40:05 & 17:33:34 \\
2023-03-19 & 08:05:14 & 10:58:42 & & & & \\
2023-03-22 & 13:01:35 & 15:55:04 & & & & \\
\hline
2024-02-21 & 10:24:32 & 13:18:02 & & 2024-08-25 & 12:16:05 & 15:09:33 \\
2024-02-24 & 15:38:06 & 18:31:36 & & 2024-08-28 & 09:34:31 & 12:28:00 \\
2024-02-27 & 15:43:35 & 18:37:04 & & 2024-09-01 & 01:40:05 & 04:29:19 \\
2024-03-01 & 14:00:04 & 16:53:33 & & 2024-09-03 & 18:19:42 & 21:13:11 \\
2024-03-04 & 14:00:03 & 16:53:32 & & 2024-09-06 & 14:34:05 & 17:27:33 \\
2024-03-07 & 14:29:06 & 17:22:34 & & 2024-09-09 & 10:36:41 & 13:30:09 \\
2024-03-09 & 12:04:20 & 14:57:48 & & 2024-09-12 & 11:06:57 & 14:00:26 \\
2024-03-13 & 13:06:04 & 15:59:32 & & 2024-09-15 & 10:51:32 & 13:45:00 \\
2024-03-16 & 14:29:05 & 17:22:34 & & 2024-09-18 & 14:00:05 & 16:53:32 \\
2024-03-18 & 21:04:04 & 23:57:33 & & 2024-09-21 & 11:11:05 & 14:04:34 \\
2024-03-22 & 11:05:43 & 13:59:12 & & & & \\
\hline
2025-03-01 & 14:33:04 & 17:26:34 & & 2025-08-27 & 11:09:02 & 14:02:31 \\
2025-03-04 & 11:21:08 & 14:14:37 & & 2025-08-30 & 14:30:06 & 17:23:34 \\
2025-03-07 & 14:35:42 & 17:29:10 & & & & \\
2025-03-11 & 00:42:40 & 03:36:09 & & & & \\
2025-03-16 & 10:04:04 & 12:57:32 & & & & \\
2025-03-19 & 04:54:57 & 07:48:25 & & & & \\
2025-03-22 & 11:50:06 & 14:43:35 & & & & \\
2025-03-25 & 14:50:06 & 17:47:22 & & & & \\
2025-03-28 & 14:04:04 & 16:57:32 & & & & \\
2025-03-31 & 12:30:03 & 15:23:31 & & & & \\
\hline
\end{tabular}
\end{table}

Previous studies have investigated the polarity reversals of the polar fields in the past SCs. With the observations of polar filaments and the H$\alpha$ synoptic maps, \cite{1986BASI} found a three-fold magnetic polarity reversal in SCs 16, 19 and 20 in the north hemisphere and in SCs 12 and 14 in the south hemisphere. \cite{makarov_duration_2003} reported that the polarity reversal in the polar regions lagged sunspot minimum 5.8 ± 0.6 years in SCs 11-23, as derived from the final disappearance of polar crown filaments.
\cite{mordvinov_reversals_2014} analyzed synoptic maps of the radial field from the Kitt Peak observatory and studied the reversals of the polar fields in SCs 21–24. They found that the northern region clearly experienced multiple reversals near the pole in time-latitude diagram during SC 21, while neither the northern nor southern regions showed multiple reversals in SC 22 and SC 23. Using data covering latitudes 55$^\circ$ to 90$^\circ$ from the Wilcox Solar Observatory (WSO), \cite{pishkalo_polar_2019} reported that the northern polar reversal in SC 21 completed in early 1981, and the southern polar reversal completed in July 1981. These reversals occurred around the maximum of the smoothed sunspot number (SSN) in each hemisphere, and this study further noted that the northern polar region showed no multiple reversals in SC 21. Since polar region observations are inherently less reliable \citep{petrie_high_resolution_2017, petrie_solar_2022}, inconsistent conclusions are yielded by different datasets and study scopes. \cite{lin_high-resolution_1994} used the high-resolution magnetograms of Big Bear Solar Observatory to investigate the polarity reversal of SC 22, which began in 1986 and reached its maximum phase around 1990. They found that the magnetic polarities in the latitudinal bands of 60$^\circ$N – 70$^\circ$N reversed first, and then 60$^\circ$S – 70$^\circ$S, 70$^\circ$S – 80$^\circ$S, 70$^\circ$N – 80$^\circ$N reversed synchronously around the maximum phase of SC. In SC 23, the southern polar polarity reversal \citep{pishkalo_polar_2019} initiated earlier than the northern one but completed later, with the northern and southern reversals completed in February 2001 and September 2001, respectively. SC 24, which began in late 2008 and reached its maximum phase around early 2014, featured a three-year stagnation of the northern polar magnetic flux as it approached zero \citep{petrie_polar_2023}.
Based on observations from the Helioseismic and Magnetic Imager (HMI; \citealt{scherrer_helioseismic_2012, schou_design_2012}) on board the Solar Dynamics Observatory (SDO; \citealt{pesnell_solar_2012}), \cite{sun_polar_2015} reported that the northern and southern magnetic fields above $\pm 60^\circ$ latitude reversed polarity in November 2012 and March 2014, respectively. They also found the northern polar region underwent multiple reversals, which has been attributed to the emergence of irregular active regions \citep{2025ApJ...988..213J}. \cite{petrie_polar_2023} compared data from the Global Oscillations Network Group and SDO/HMI for SC 24, and found that variations in magnetic flux from both datasets followed the same triple reversal pattern. \cite{yang_long-term_2024} analyzed polar photospheric magnetic field data released by Nagoya University from the Spectropolarimeter (SP; \citealt{lites_hinode_2013}) of Solar Optical Telescope (SOT; \citealt{tsuneta_solar_2008}) on board the Hinode \citep{kosugi_hinode_2007} satellite. They revealed that the magnetic field polarity in each solar polar region reversed sequentially from 70$^\circ$ latitude toward the pole during the maximum phase of SC 24. They also demonstrated that the magnetic flux density remained close to zero for approximately three years around the time of the northern polar reversal.

SC 25 began in December 2019 and has now passed its solar maximum around 2024, entering the decline phase \footnote{\url{https://www.sidc.be/SILSO/monthlyssnplot}}. Predictions and observations of the polarity reversal times of SC 25 have already been conducted. \cite{jha_predicting_2024} employed the advective flux transport model to derive predictions for the polarity reversal times of SC 25 above $60^\circ$ latitude in both hemispheres. Their results indicate that the northern polar region reversed between June and November 2024, while the southern polar region is expected to reverse between November 2024 and July 2025. \cite{bilenko_solar_2026} utilized polar photospheric magnetic field data from WSO to investigate polarity reversals in $45^\circ-70^\circ$ of both the northern and southern hemispheres across SC 21–25, and reported that the polarity reversal of SC 25 initiated in Carrington Rotation (CR) 2269 and completed in CR 2280, corresponding to the period between March 2023 and February 2024.

Since SC 25 has only recently passed its solar maximum, observational studies remain scarce, and there is thus a lack of detailed investigations of polar magnetic fields. In this paper, we utilize the Hinode/SP data to construct polar composite maps of radial magnetic flux density during polarity reversal, compare polar cap magnetic fields with hemispheric sunspot numbers, and analyze SC 25’s polar magnetic polarity reversal across different latitudinal ranges.

\section{Observations and Data analysis}
\begin{figure}[pt]
\centerline{\includegraphics[width=1\textwidth,clip=]{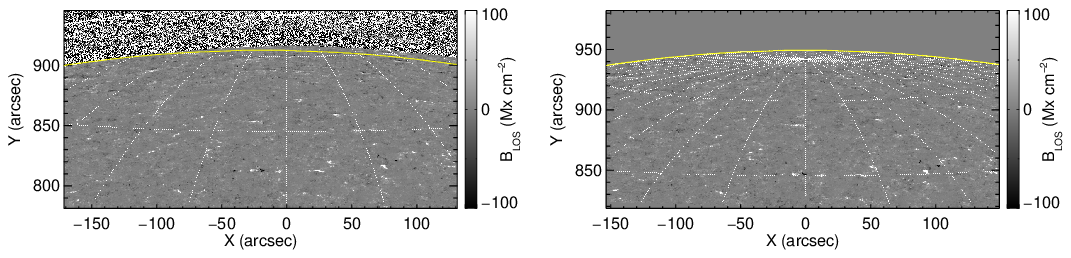}}
\caption{Maps of line-of-sight magnetic flux density before (left) and after (right) the calibration. White dotted curves denote latitude and longitude grid with $10^\circ$ intervals. Yellow curves indicate the correct positions of the solar limb.}
\label{fig:fig1}
\end{figure}

\begin{figure}
\centerline{\includegraphics[width=0.75\textwidth,clip=]{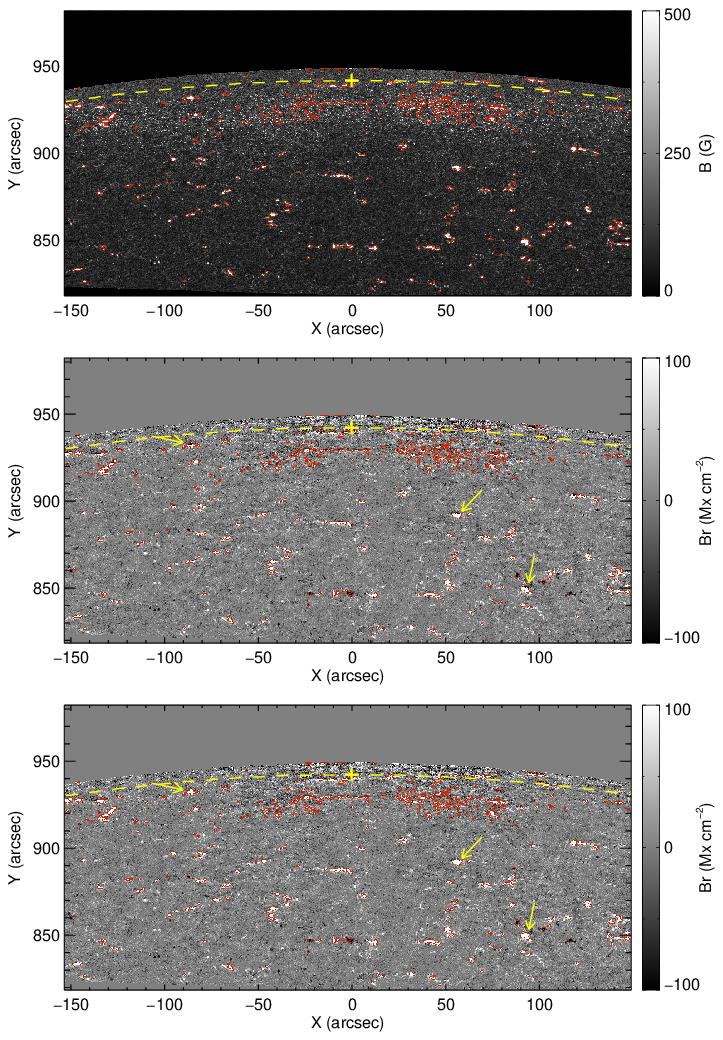}}
\caption{Maps of intrinsic total magnetic field strength (top panel) and radial magnetic flux density after resolving the 180$^\circ$ ambiguity (middle panel) and after subsequent removal of spurious polarities (bottom panel). Red curves denote the boundaries of magnetic patches, where the intrinsic total magnetic field strength exceeds 200 G across the entire patch.Yellow plus signs indicate poles and yellow curves represent $\pm 90^\circ$ longitude. Yellow arrows show a few examples of regions with spurious polarity.}
\label{fig:fig2}
\end{figure}

The solar rotation axis is inclined at $7.25^\circ$ relative to the ecliptic plane, making September the best time for observing the northern polar region and March for the southern polar region \citep{petrie_solar_2015}. While previous studies \citep{yang_long-term_2024,yang_meridional_2024} adopted SP data from 2012 to 2021 to investigate SC 24, we use Level 2 data (DOI: 10.5065/D6JH3J8D) from SOT/SP on board Hinode, acquired near March and September each year spanning 2022 to 2025, to study the polar field reversal of SC 25. 
SP scans solar polar regions to obtain line profiles of two magnetically sensitive lines (Fe I \SI{630.25} nm and \SI{630.15} nm; \citealt{tsuneta_solar_2008}). SP Level 2 data include vector magnetograms and a set of associated variables, and they are generated by the High Altitude Observatory team via the Milne-Eddington gRid Linear Inversion Network \citep{lites_b_suite_2007} with Level 1 calibrated spectra \citep{lites_sp_prep_2013} serving as the input. The data can be publicly available through the Lockheed Martin Solar and Astrophysics Laboratory SOT Data Center \footnote{\url{https://sot.lmsal.com/data/sot/level2hao/}}. The specific datasets employed herein are presented in Table \ref{tabel1}. The pixel scale of data is 0.3$''$ along the SP scanning direction and 0.32$''$ along the SP slit.

There is a mismatch between the SP-derived coordinates and the observed image, see the left panel of Figure \ref{fig:fig1}, which is caused by various factors \citep{tsuneta_solar_2008}. We extract a sub field of view from the HMI magnetogram corresponding to the field of view of the SP magnetogram and align them with characteristic features (e.g. strong magnetic patches) near the central meridian, and then we derive the correct heliographic position. Since SP data are acquired by slit scanning, the solar limb in the magnetogram still shows a small discrepancy from the corrected position (yellow curves in Figure \ref{fig:fig1}). We shift the image upward or downward column by column, aligning the solar limb completely. The calibrated magnetogram is shown in the right panel of Figure \ref{fig:fig1}. HMI data can be obtained from the Joint Science Operations Center \footnote{\url{http://jsoc.stanford.edu/ajax/exportdata.html}}.

In the observed magnetograms, there is an inherent $180^\circ$ ambiguity in the azimuth, i.e., there are two solutions for the azimuth that differ by $180^\circ$. 
Our disambiguation method follows that of \cite{ito_is_2010}. This method relies on the physical assumption that the magnetic field is either nearly horizontal or nearly vertical (or undetermined) with respect to the local surface. The assumption is based on the bimodal distribution reported by \cite{orozco_suarez_quiet-sun_2007} using quiet-sun vector magnetic field data, with peaks at the vertical and horizontal directions. This method has been widely adopted and validated in studies of polar magnetic fields with Hinode and Solar Orbiter observations (e.g., \citealt{shiota_polar_2012,petrie_high_resolution_2017,kubo_comparison_2025,yang_variations_2025,calchetti_first_2025}). Based on the inclination angle $\gamma$ between the magnetic field vector and the local normal direction, these solutions are classified into three categories: vertical fields ($0^\circ \leq \gamma \leq 40^\circ$ or $140^\circ \leq \gamma \leq 180^\circ$), horizontal fields ($70^\circ \leq \gamma \leq 110^\circ$), and other fields that are not further analyzed. If both solutions correspond to vertical fields, the one closer to the local normal is selected as true solution. If one of solutions is either vertical or horizontal and the other is non-vertical/non-horizontal, the former is preferred. All remaining pixels are discarded. After resolving the $180^\circ$ ambiguity, the radial magnetic flux density is calculated \citep{kubo_comparison_2025} using the formula $B_r = f \cdot B \cdot \cos(\gamma)$, where $f$ is magnetic filling factor, $B$ is the intrinsic total magnetic field strength, $\gamma$ is the angle between the magnetic field vector and the local normal.

Previous studies \citep{tsuneta_magnetic_2008,yang_variations_2025} have shown that patchy magnetic islands with strong strengths possess fanning-out magnetic structures. These structures should possess the same polarity, while some exhibit spurious polarity due to the projection effect \citep{deng_vector_1999} in polar radial magnetic flux density maps, as marked by the yellow arrows in the middle panel of Figure \ref{fig:fig2}. \cite{ito_is_2010} noted that the azimuth at the locations of spurious polarity can be manually corrected via visual inspection. We rectify such spurious polarities through a thresholding process. For each magnetogram, continuous areas with more than 10 pixels and where the intrinsic total magnetic field strength exceeds 200 G across the entire area are defined as strong magnetic patches. As we focus on the net flux in the polar regions, it is unnecessary to set an signal-to-noise threshold. Selected magnetic patches are marked with red contours, and each magnetic patch should have a single polarity. On the other hand, region of one polarity in the patch closer to the limb is less reliable, we intend to regard limb-side polarity as spurious. The spurious-polarity area is generally smaller than true-polarity area, i.e., the ratio of the spurious-polarity area to true-polarity area is less than 100\%, but it could exceed 100\% for the regions very close to the limb, i.e., beyond $\pm$90$^\circ$ longitude (above yellow curves in Figure \ref{fig:fig2}) due to large noise and foreshortening effect. Based on visual inspection, a threshold of 150\% is the most suitable for determining spurious polarity. Besides, a higher threshold makes misidentification more likely and 150\% is conservative. In the spurious polarity region, the $B_r$ values are reversed. The result of this correction is shown in the bottom panel of Figure \ref{fig:fig2}. To improve data reliability further, we only include regions within $\pm$90$^\circ$ longitude in the subsequent analysis, where the ratio of the spurious-polarity area to true-polarity area is less than 100\%.

\section{Results} \label{sec:Results}
\begin{figure}[t]
\centering
\includegraphics[width=1\textwidth]{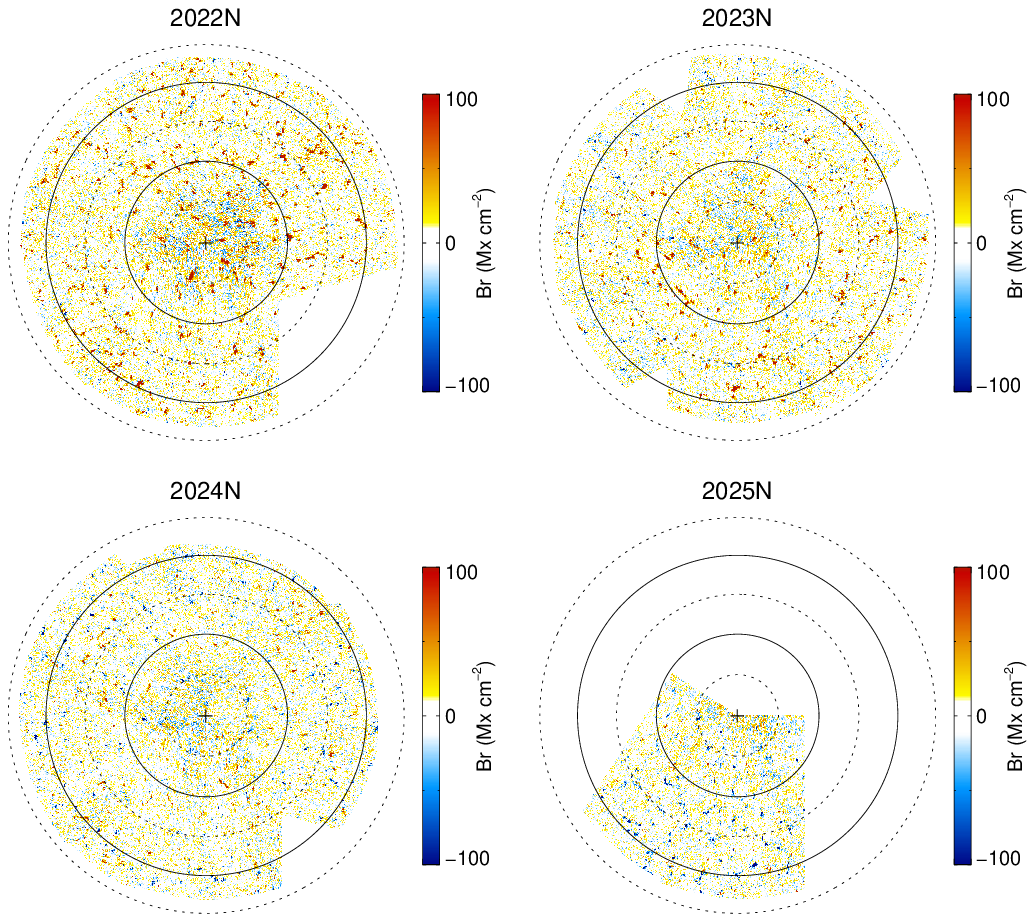}
\caption{Northern polar composite maps of radial magnetic flux density from 2022 to 2025. Plus signs (+) denote the poles, black solid curves correspond to $10^\circ$ latitudinal intervals, and black dotted curves to $5^\circ$ intervals.}
\label{topView_n}
\end{figure}

\begin{figure}[t]
\centering
\includegraphics[width=1\textwidth]{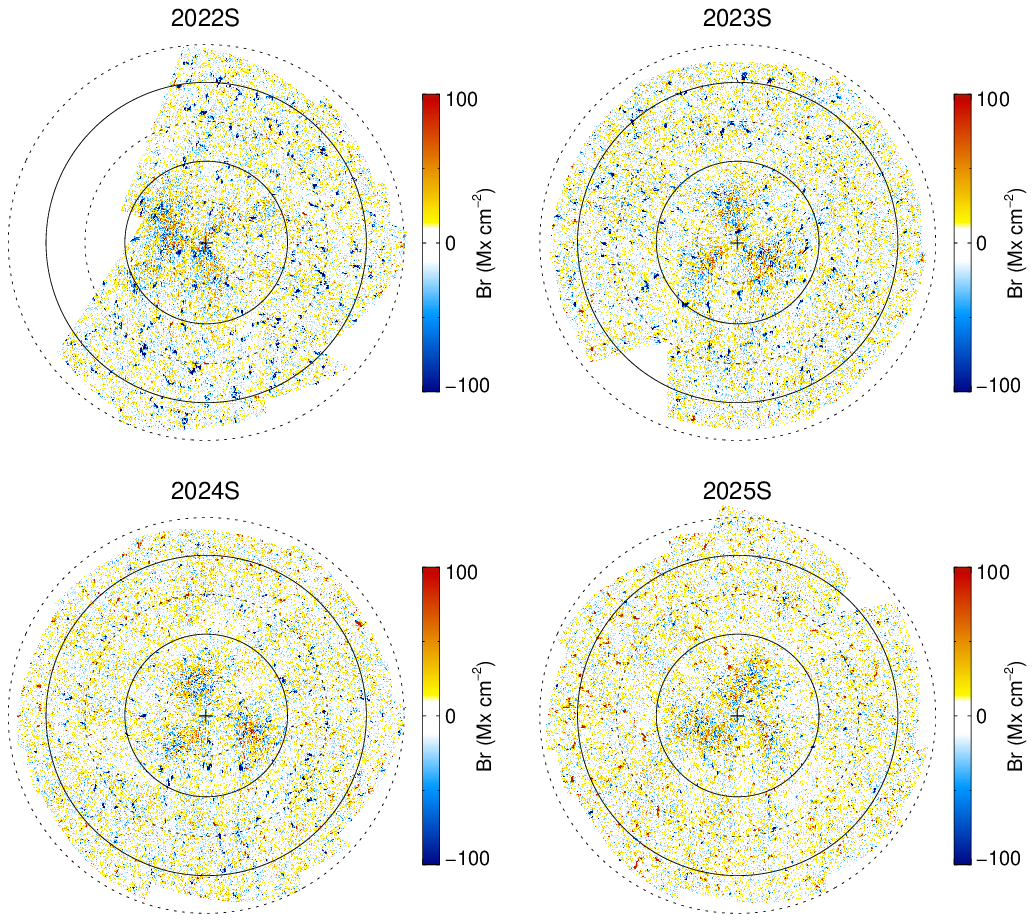}
\caption{Southern polar composite maps of radial magnetic flux density from 2022 to 2025. Plus signs (+) denote the poles, black solid curves correspond to $10^\circ$ latitudinal intervals, and black dotted curves to $5^\circ$ intervals.}
\label{topView_s}
\end{figure}

We perform geometric transformations on each magnetogram to convert the Hinode view to a solar polar top-down view \citep{yang_long-term_2024, yang_meridional_2024}, hereafter referred to as polar view. Following the solar differential rotation formula proposed by \cite{howard_solar_1990}, a rotation period (34.14 days) is determined with $80^\circ$ as the benchmark for image rotation. Based on the timestamp of each magnetogram, its position relative to the others can be determined, then the magnetograms are stitched in sequence. Therefore, one polar-view composite map is generated annually for the northern (Figure \ref{topView_n}) and the southern polar regions (Figure \ref{topView_s}), respectively.

In Figure \ref{topView_n}, the positive-polarity magnetic field was predominant in the northern polar cap in September 2022, with a small number of negative-polarity magnetic patches mainly confined to latitudes below 70$^\circ$. In September 2023, the number of negative-polarity magnetic patches increased and they were primarily distributed at latitudes below $80^\circ$ while positive-polarity magnetic flux remained predominant across the entire polar cap. In September 2024, both positive- and negative-polarity magnetic patches appeared in all latitudes and it became difficult to distinguish which polarity was predominant. From 2022 to 2024, the polar magnetic field has been gradually weakening. By September 2025, as evidenced by the limited portion of the available polar composite map, the negative-polarity magnetic field had already become predominant, which means the northern polarity reversed between September 2024 and September 2025. As shown in Figure \ref{topView_s}, the southern polar region exhibited an opposite trend. In March 2022, the polar magnetic field was almost entirely of negative polarity. The strong positive-polarity magnetic patches appeared in the $65^\circ-70^\circ$ latitude range in March 2023 and appeared at $75^\circ$ latitude in March 2024, showing an overall trend of advancing toward the pole. The polar magnetic field strength gradually weakened as well. By March 2025, the entire southern polar polarity has reversed to being dominated by positive polarity. It also means that the southern polarity reversed between March 2024 and March 2025.

\begin{figure}[t]
	\centering
	\includegraphics[width=0.7\textwidth]{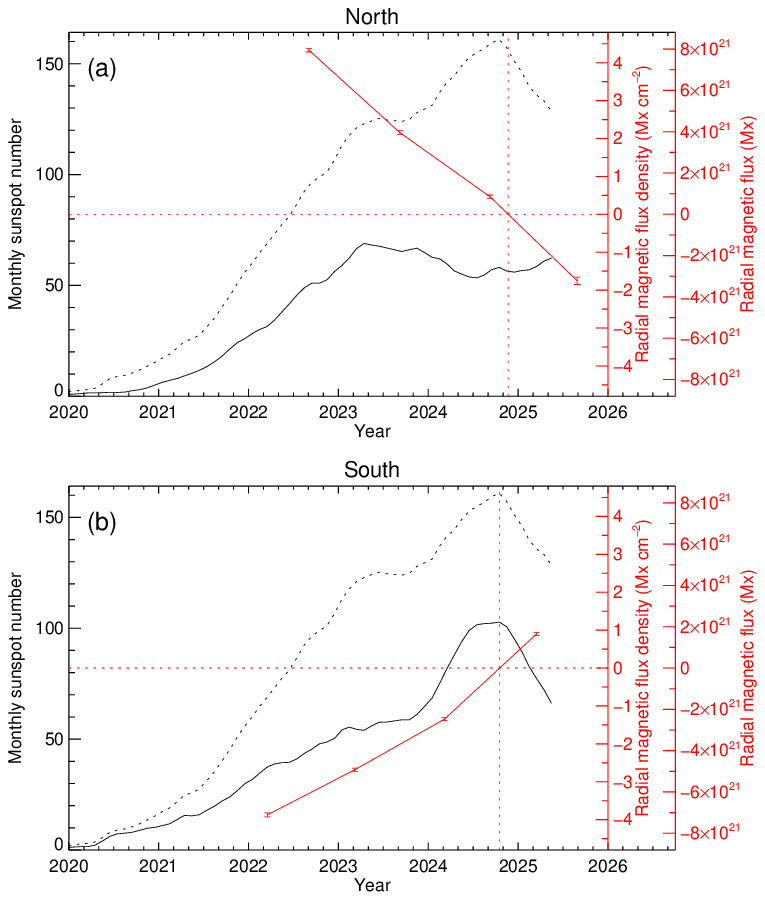}
	\caption{Evolution of 13-month smoothed monthly hemispheric sunspot number and the average radial magnetic flux density and the radial magnetic flux with the latitudinal range from $70^\circ$ N to $90^\circ$ N (panel a) and from $70^\circ$ S to $90^\circ$ S (panel b). The black dotted curves show the evolution of total sunspot number; the black solid curve in panels (a) and (b) denotes the northern and southern hemispheric sunspot number, respectively. Red solid curves represent radial magnetic flux density and radial magnetic flux. Red vertical dotted curves correspond to the estimated polar polarity reversal times. Error bars indicate the values propagated from the errors of the inversion parameters.}
	\label{fig:sunspots}
\end{figure}

To investigate the relationship between polar field evolution and solar activity during SC 25, we further adopt 13-month smoothed monthly hemispheric sunspot number data \citep{https://doi.org/10.24414/qnza-ac80,clette_recalibration_2023} provided by the Sunspot Index and Long-term Solar Observations (SILSO), and the polar cap is corresponds to the latitudinal interval $70^\circ - 90^\circ$. The average radial magnetic flux density and the radial magnetic flux are calculated as $\bar{B}_{r_{k}} = \sum_{i,j} B_{r_{ijk}} S_{ijk} / \sum_{i,j} S_{ijk}$, and $\Phi_{k} = \bar{B}_{r_{k}} A$, where ${B}_{r_{ijk}}$ is radial magnetic flux density and $S_{ijk}$ is corresponding spherical area of the $i$-th pixel with a longitude range of $-90^\circ$ to $90^\circ$ on the $j$-th magnetogram in year $k$, and $A$ denotes the area of the spherical cap within the latitudinal range $\pm \left[70^\circ - 90^\circ \right]$ with radius $6.96 \times 10^{10}$ cm. The errors of the average radial magnetic flux density and radial magnetic flux are calculated as $\sigma_{Br_{k}} = \sqrt{\sum_{i,j} (\sigma_{Br_{ijk}} S_{ijk} / \sum_{i,j} S_{ijk})^2}$, and $\sigma_{\Phi_{k}} = \sigma_{Br_{k}} A$, respectively. $\sigma_{Br_{ijk}}$ is the value propagated from the errors of the inversion parameters, and the subscripts $i$, $j$, $k$ follow the same definition as above. Measurement data are sampled at one point per year and linearly interpolated between adjacent points. We determine the polar field reversal times from this interpolated time series.

The northern hemispheric sunspot number reached its maximum (SSN$=$68.9) in April 2023 (Figure \ref{fig:sunspots} (a)). The northern polar polarity reversal presumably occurred in November 2024. Around the polarity reversal, $\bar{B}_{r}$ changed from 0.47 Mx cm$^{-2}$ in September 2024 to -1.75 Mx cm$^{-2}$ in September 2025, and $\Phi$ changed from $0.86 \times 10^{21}$ Mx to $-3.21 \times 10^{21}$ Mx over the same period. From 2022 to 2025, the average decrease rate of the radial magnetic flux density and magnetic flux in the northern polar cap are approximately $1.99$ Mx cm$^{-2}$ yr$^{-1}$ and $3.66 \times 10^{21}$ Mx yr$^{-1}$, respectively.
The southern hemispheric sunspot number reached its maximum (SSN$=$102.8) in October 2024 (Figure \ref{fig:sunspots} (b)). The southern polar polarity reversal possibly occurred in October 2024. Around the polarity reversal, $\bar{B}_{r}$ changed from -1.35 Mx cm$^{-2}$ in March 2024 to 0.90 Mx cm$^{-2}$ in March 2025, while $\Phi$ changed from $-2.47 \times 10^{21}$ Mx to $1.65 \times 10^{21}$ Mx. The average increase rate of the radial magnetic flux density and magnetic flux in the southern polar cap from 2022 to 2025 are approximately $1.57$ Mx cm$^{-2}$ yr$^{-1}$ and $2.87 \times 10^{21}$Mx yr$^{-1}$, respectively.

\begin{figure}[t]
	\centering
	\includegraphics[width=\textwidth]{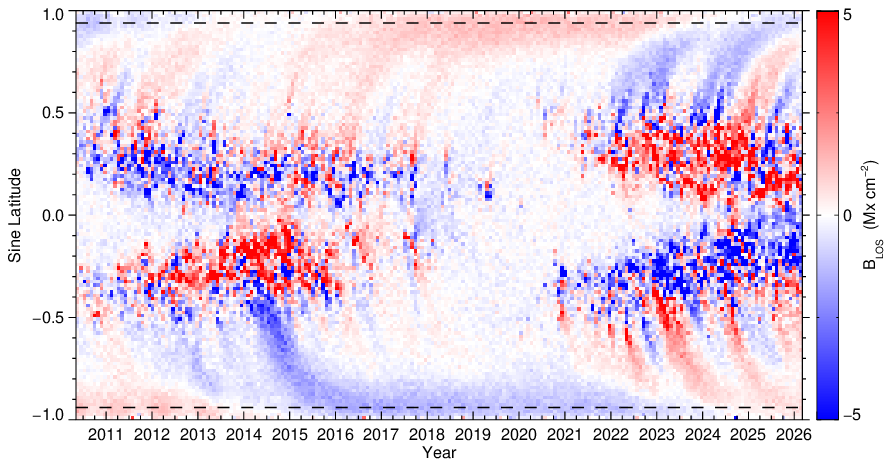}
	\caption{Latitude–time plot of the line-of-sight magnetic field from HMI synoptic maps. The black dashed lines indicate the latitudes of $\pm 70^\circ$.}
	\label{fig:timelat}
\end{figure}

\begin{figure}[t]
	\centering
	\includegraphics[width=0.8\textwidth]{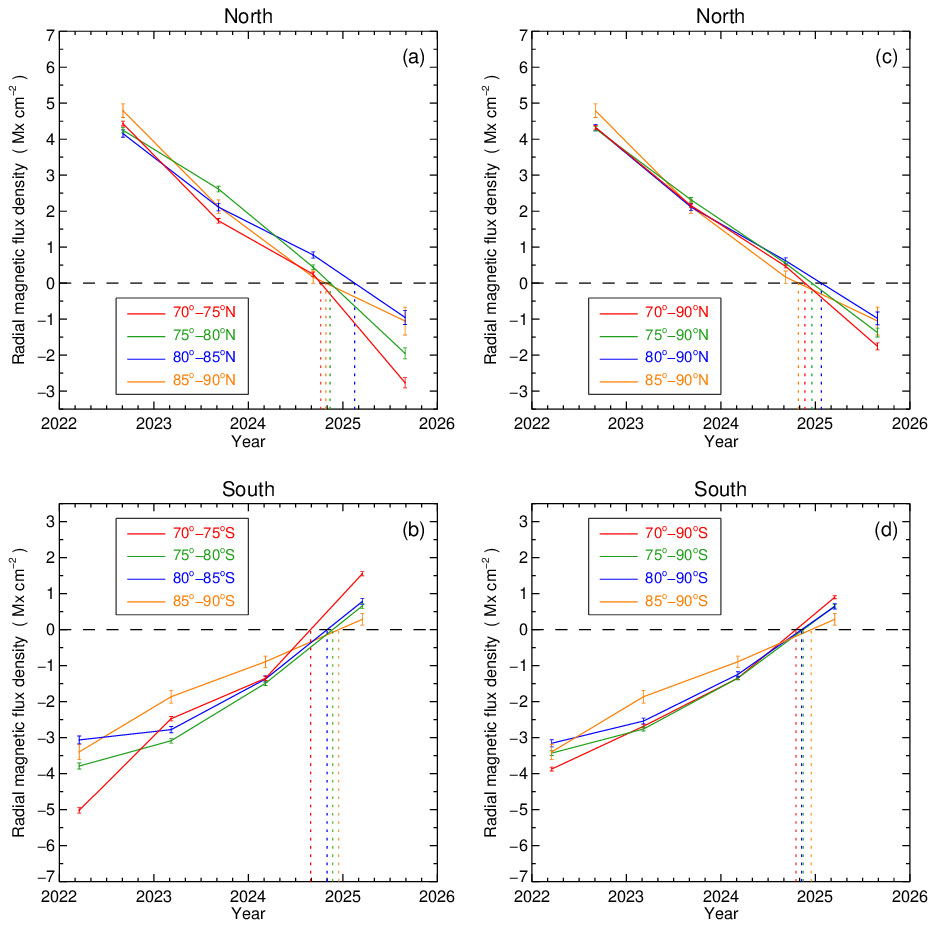}
	\caption{Evolution of radial magnetic flux density in the northern (top panels) and southern (bottom panels) polar regions. Error bars indicate the values propagated from the errors of the inversion parameters.}
	\label{fig:Br_lat_cap}
\end{figure}

\begin{figure}[htbp]
	\centerline{\includegraphics[width=0.8\textwidth,clip=]{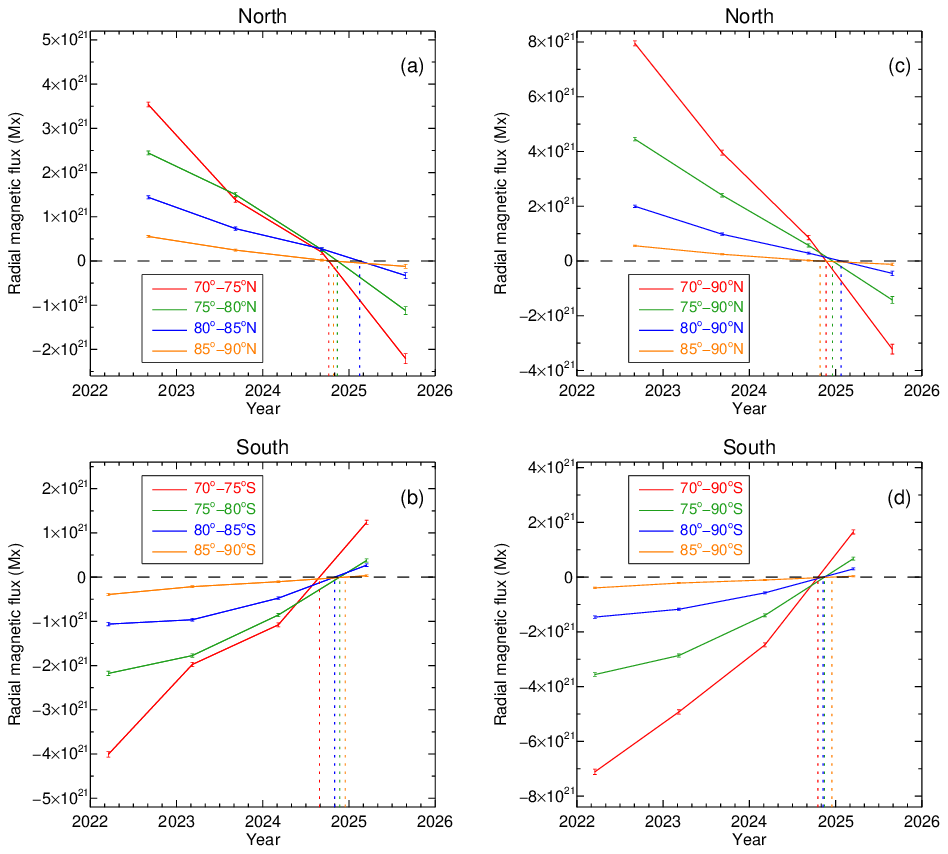}}
	\caption{Evolution of radial magnetic flux in the northern (top panels) and southern (bottom panels) polar regions.Error bars indicate the values propagated from the errors of the inversion parameters.}
	\label{fig:flux_lat_cap}
\end{figure}

To further validate the reliability of reversal times, we use HMI synoptic line-of-sight magnetic field maps \citep{liu_comparison_2012} to construct a latitude-time plot. One synoptic line-of-sight magnetic field map is available per CR and we average each map in longitude and arrange them sequentially in time, as shown in Figure \ref{fig:timelat}. In the latitude–time plot, we find that both polar regions (above $\pm 70^\circ$) may have undergone polarity reversal in 2024. Besides, the latitude-time plot also shows that magnetic flux is transported from low latitudes toward the polar regions.

We also calculate the radial magnetic flux density and radial magnetic flux for different latitudinal bands following the formula used earlier for the polar cap. Figures \ref{fig:Br_lat_cap} and \ref{fig:flux_lat_cap} show the evolution of radial magnetic flux density and radial magnetic flux, respectively, for different latitudinal band and polar cap ranges.At a 1-year cadence, each of the eight panels shows a monotonic variation without multiple reversals. The polarity reversal times of the northern and southern polar regions are listed in Table \ref{table2}. 
In the northern polar region, the polarity first reversed in the $70^\circ - 75^\circ$ latitudinal band around October 2024 and the latest reversal may occurred in the $80^\circ - 85^\circ$ latitudinal band around February 2025, lasting approximately 4 months. The $70^\circ - 75^\circ$, $75^\circ - 80^\circ$ and $80^\circ - 85^\circ$ latitudinal bands reversed in sequence.
In the southern polar region, the polarity first reversed in the $70^\circ - 75^\circ$ latitudinal band around August 2024, then in the $80^\circ - 85^\circ$, slightly later in the $75^\circ - 80^\circ$, and the latest in the $85^\circ - 90^\circ$ around December 2024, lasting approximately 4 months. During 2022–2025, the predominant-polarity radial magnetic flux density was the highest in 2022 for all latitudinal bands (Figures \ref{fig:Br_lat_cap} (a) and (c)), ranging from 4 to 5 Mx cm$^{-2}$ in the northern polar region and 3 to 5 Mx cm$^{-2}$ in the southern polar region. In September 2024, shortly before the northern polar region’s polarity reversal, the radial magnetic flux density for all latitudinal bands was only $\sim 1$ Mx cm$^{-2}$, as shown in Figure \ref{fig:Br_lat_cap} (a).
Notably, in this work, the completion of the polar field reversal is identified as the moment when the average magnetic flux density over the polar cap above $70^\circ$ latitude ($70-90^\circ$) changed to the opposite sign, and thus the polarity reversals at the north and south polar regions are found to be completed in November 2024 and October 2024, respectively.

\begin{table}
\caption{Estimated polarity reversal times of different regions}
\begin{tabular*}{0.8\linewidth}{@{\extracolsep{\fill}} lccc}
\hline
 & Range & North & South\\
\hline
\multirow{4}{*}{Latitudinal Bands} & $70^\circ - 75^\circ$ & 2024-10 & 2024-08\\
 & $75^\circ - 80^\circ$ & 2024-11 & 2024-11\\
 & $80^\circ - 85^\circ$ & 2025-02 & 2024-11\\
 & $85^\circ - 90^\circ$ & 2024-10 & 2024-12\\
\hline
\multirow{4}{*}{Polar Caps} & $70^\circ - 90^\circ$ & 2024-11 & 2024-10\\
 & $75^\circ - 90^\circ$ & 2024-12 & 2024-11\\
 & $80^\circ - 90^\circ$ & 2025-01 & 2024-11\\
 & $85^\circ - 90^\circ$ & 2024-10 & 2024-12\\
\hline
\end{tabular*}
\label{table2}
\end{table}

\section{Conclusions and Discussion} \label{sec:CD}
The polar composite maps reveal the year-to-year variations in the predominant polarity, with the northern polar region gradually transitioning from positive to negative, and the opposite trend observed in the southern polar region. By geometrically transforming Hinode-view radial magnetic flux density maps to the polar view and stitching, we obtained polar-view composite maps. The composite maps of the northern polar region show that the positive polarity weakened over the years, until there was no predominant polarity in 2024. The southern polar region was dominated by negative polarity, which weakened over the years 2022-2024 and reversed to positive-polarity predominance in 2025.

The northern polar cap above $70^\circ$ is estimated to have experienced polarity reversal around November 2024, with the event lagging the hemispheric sunspot number maximum by approximately 19 months, and the southern polar cap is estimated to have experienced polarity reversal around October 2024, almost coinciding with the hemispheric sunspot number maximum. Over the period 2022–2025, the radial magnetic flux in the northern and southern polar caps exhibited average rates of change of $-3.66 \times 10^{21}$ Mx yr$^{-1}$ and $2.87 \times 10^{21}$ Mx yr$^{-1}$, respectively. In the southern polar region, although the polar reversal time coincides with hemispheric sunspot maximum time, the trailing-polarity flux continues to be transported toward the poles afterward, causing the polar magnetic field to further strengthen. The latitude-time plot shows that the polar field reversal at both poles occurred in 2024, which are generally consistent with those from the Hinode data, confirming the robustness of our results. \cite{yang_long-term_2024} compared the radial magnetic flux density and hemispheric sunspot numbers before and after the polarity reversal of SC 24. Their results showed that the northern polar cap polarity reversal occurred one and a half years after the northern hemispheric sunspot number peaks, while the southern polar cap polarity reversal took place four months after the southern hemispheric sunspot number reached its maximum. The corresponding time lags in our study follow the similar pattern. \cite{pishkalo_polar_2019} reported that reversals of polar magnetic field in SC 21-24 were completed first in the northern hemisphere. Previous studies \citep{sun_polar_2015,janardhan_solar_2018,yang_long-term_2024} on SC 24’s polar cap polarity reversal showed that the southern reversal occurred approximately one and a half years later than the northern, whereas our results show the southern reversal precedes its northern by approximately only one month in SC 25.

Polar magnetic polarity reversal of SC 25 tends to exhibit a sequential progression from lower latitudes toward the poles. In the northern polar region, the $70^\circ-75^\circ$ latitudinal band reversed the earliest around October 2024 and the $80^\circ-85^\circ$ latitudinal band reversed the latest around February 2025. Reversals occurred sequentially from lower latitudinal regions to higher latitudinal regions except for $85^\circ-90^\circ$ latitudinal band. Since magnetograms of the northern polar region for 2025 have not yet been fully released, we can only rely on the two existing magnetograms for calculations. Additionally, the $85^\circ-90^\circ$ latitudinal band has a small area with few data, possibly making it an exception. In the southern polar region, there is also a trend that lower latitudinal bands reverse earlier and higher latitudinal bands reverse later. The $70^\circ-75^\circ$ latitudinal band reversed the earliest around August 2024 and the $85^\circ-90^\circ$ latitudinal band reversed the latest around December 2024. Notably, the $80^\circ-85^\circ$ latitudinal band reversed slightly earlier than the $75^\circ-80^\circ$ latitudinal band. In the absence of more detailed data around the polarity reversal dates, the actual order of the two closely reversal times needs further validation. \cite{yang_long-term_2024}'s results indicate that during SC 24, the latitudinal bands between $70^\circ$ and $90^\circ$ at $5^\circ$ intervals in the polar regions of both hemispheres reversed sequentially from the $70^\circ-75^\circ$ band to the $85^\circ-90^\circ$ band. The polarity reversal trend indicated by our results is almost consistent with theirs. 
Such sequentially reversed polar fields can be naturally explained by the flux transport model \citep{van_ballegooijen_magnetic_1998,jiang_magnetic_2014,wang_surface_2017}. Active regions emerge at low latitudes in accordance with Joy's law, and their trailing-polarity magnetic flux is transported toward the polar regions by the meridional flow, driving the polar field reversal and producing a sequential reversal from lower to higher latitudes within each polar cap. Following the reversal, the poleward meridional flow together with fluctuations in active region emergence produces prominent poleward flux surges and variations in the polar field, with trailing polarities ultimately dominating in each hemisphere. This pattern can be clearly observed in the latitude–time plot (Figure \ref{fig:timelat}). In addition, although 1-year cadence yields a monotonic variation with no multiple reversals, shorter-timescale polarity reversals may exist due to large short-term fluctuations.

\cite{bilenko_solar_2026} reported that the polarity reversal in the $45^\circ-70^\circ$ latitudinal bands of the northern and southern hemispheres during SC 25 began in March 2023 and completed in February 2024.
\cite{jha_predicting_2024} predicted the polarity reversal times for SC 25, with the northern polar region ($60^\circ-90^\circ$) reversing between June 2024 and November 2024, and the southern polar region ($60^\circ-90^\circ$) reversing between November 2024 and July 2025.
While the latitudinal range of our study is $70^\circ-90^\circ$, based on the trend that lower-latitude regions reverse polarity first, the polarity reversal time for the $45^\circ-70^\circ$ and $60^\circ-90^\circ$ ranges should be earlier than that for the $70^\circ-90^\circ$, i.e., prior to November 2024 in the northern hemisphere and October 2024 in the southern hemisphere. Therefore, our results are compatible with Bilenko's results in both hemispheres and with Jha \& Upton’s predicted reversal time in the northern hemisphere, whereas our southern polarity reversal time precedes theirs.

\begin{acks}
This research is supported by Beijing Natural Science Foundation (1252034), the National Key R\&D Programs of China (2022YFF0503800, 2022YFF0503003), the National Natural Science Foundations of China (12573056, 12533010, 12350004), the Strategic Priority Research Programs of the Chinese Academy of Sciences (XDB0560000), the Youth Innovation Promotion Association CAS, the Specialized Research Fund for State Key Laboratory of Solar Activity and Space Weather, China's Space Origins Exploration Program, and Key Innovation Team of China Meteorological Administration Space Weather Monitoring and Alerting (CMA2024ZD01). Hinode is a Japanese mission developed and launched by ISAS/JAXA, with NAOJ as domestic partner and NASA and STFC (UK) as international partners. It is operated by these agencies in co-operation with ESA and NSC (Norway). Hinode SOT/SP Inversions were conducted at NCAR under the framework of the Community Spectropolarimetric Analysis Center (CSAC; http://www2.hao.ucar.edu/csac).The smoothed monthly hemispheric sunspot number data are from the World Data Center SILSO, Royal Observatory of Belgium, Brussels.
\end{acks}

%
%
%
%
%
%
%

\bibliographystyle{spr-mp-sola}
\bibliography{pr25}  

@misc{https://doi.org/10.24414/qnza-ac80,
doi = {10.24414/QNZA-AC80},
url = {https://www.astro.oma.be/doi/ROB-SIDC-SILSO_SunspotNumberV2.html},
author = {Clette, Frédéric and Lefèvre, Laure},
title = {SILSO Sunspot Number V2.0},
publisher = {World Data Center SILSO, Royal Observatory of Belgium (ROB)},
year = {2015}
}

@ARTICLE{2025ApJ...988..213J,
       author = {{Jin}, Chunlan and {Yang}, Shuhong and {Zhou}, Guiping and {Hu}, Jialiang and {Wang}, Jingxiu},
        title = "{A Possible Source of Multiple Polarity Reversals of the Polar Magnetic Field during the Maximum Phase of Solar Cycle}",
      journal = {APJ},
         year = 2025,
        month = aug,
       volume = {988},
       number = {2},
          eid = {213},
        pages = {213},
          doi = {10.3847/1538-4357/ade681},
       adsurl = {https://ui.adsabs.harvard.edu/abs/2025ApJ...988..213J}
}

@article{babcock_suns_1959,
	title = {The {Sun}'s {Polar} {Magnetic} {Field}.},
	volume = {130},
	issn = {0004-637X, 1538-4357},
	url = {http://adsabs.harvard.edu/doi/10.1086/146726},
	doi = {10.1086/146726},
	language = {en},
	urldate = {2025-12-23},
	journal = {The Astrophysical Journal},
	author = {Babcock, Harold D.},
	month = sep,
	year = {1959},
	pages = {364},
}

@article{shiota_polar_2012,
	title = {{POLAR} {FIELD} {REVERSAL} {OBSERVATIONS} {WITH} \textit{{HINODE}}},
	volume = {753},
	issn = {0004-637X, 1538-4357},
	url = {https://iopscience.iop.org/article/10.1088/0004-637X/753/2/157},
	doi = {10.1088/0004-637X/753/2/157},
	language = {en},
	number = {2},
	urldate = {2025-11-25},
	journal = {ApJ},
	author = {Shiota, D. and Tsuneta, S. and Shimojo, M. and Sako, N. and Orozco Suárez, D. and Ishikawa, R.},
	month = jul,
	year = {2012},
	pages = {157},
}

@article{kosugi_hinode_2007,
	title = {The {Hinode} ({Solar}-{B}) {Mission}: {An} {Overview}},
	volume = {243},
	copyright = {http://www.springer.com/tdm},
	issn = {0038-0938, 1573-093X},
	shorttitle = {The {Hinode} ({Solar}-{B}) {Mission}},
	url = {http://link.springer.com/10.1007/s11207-007-9014-6},
	doi = {10.1007/s11207-007-9014-6},
	language = {en},
	number = {1},
	urldate = {2025-11-25},
	journal = {Sol Phys},
	author = {Kosugi, T. and Matsuzaki, K. and Sakao, T. and Shimizu, T. and Sone, Y. and Tachikawa, S. and Hashimoto, T. and Minesugi, K. and Ohnishi, A. and Yamada, T. and Tsuneta, S. and Hara, H. and Ichimoto, K. and Suematsu, Y. and Shimojo, M. and Watanabe, T. and Shimada, S. and Davis, J. M. and Hill, L. D. and Owens, J. K. and Title, A. M. and Culhane, J. L. and Harra, L. K. and Doschek, G. A. and Golub, L.},
	month = sep,
	year = {2007},
	pages = {3--17},
}

@article{mordvinov_reversals_2014,
	title = {Reversals of the {Sun}’s {Polar} {Magnetic} {Fields} in {Relation} to {Activity} {Complexes} and {Coronal} {Holes}},
	volume = {289},
	copyright = {http://www.springer.com/tdm},
	issn = {0038-0938, 1573-093X},
	url = {http://link.springer.com/10.1007/s11207-013-0456-8},
	doi = {10.1007/s11207-013-0456-8},
	language = {en},
	number = {6},
	urldate = {2025-11-19},
	journal = {Sol Phys},
	author = {Mordvinov, A. V. and Yazev, S. A.},
	month = jun,
	year = {2014},
	pages = {1971--1981},
}

@article{charbonneau_dynamo_2020,
	title = {Dynamo models of the solar cycle},
	volume = {17},
	issn = {2367-3648, 1614-4961},
	url = {https://link.springer.com/10.1007/s41116-020-00025-6},
	doi = {10.1007/s41116-020-00025-6},
	language = {en},
	number = {1},
	urldate = {2025-11-19},
	journal = {Living Rev Sol Phys},
	author = {Charbonneau, Paul},
	month = {12},
	year = {2020},
	pages = {4},
}

@article{jha_predicting_2024,
	title = {Predicting the {Timing} of the {Solar} {Cycle} 25 {Polar} {Field} {Reversal}},
	volume = {962},
	issn = {2041-8205, 2041-8213},
	url = {https://iopscience.iop.org/article/10.3847/2041-8213/ad20d2},
	doi = {10.3847/2041-8213/ad20d2},
	language = {en},
	number = {1},
	urldate = {2025-11-19},
	journal = {ApJL},
	author = {Jha, Bibhuti Kumar and Upton, Lisa A.},
	month = feb,
	year = {2024},
	pages = {L15},
}

@article{petrie_high_resolution_2017,
	title = {High-{Resolution} {Vector} {Magnetograms} of the {Sun}’s {Poles} from {Hinode}: {Flux} {Distributions} and {Global} {Coronal} {Modeling}},
	volume = {292},
	issn = {0038-0938, 1573-093X},
	shorttitle = {High-{Resolution} {Vector} {Magnetograms} of the {Sun}’s {Poles} from {Hinode}},
	url = {http://link.springer.com/10.1007/s11207-016-1034-7},
	doi = {10.1007/s11207-016-1034-7},
	language = {en},
	number = {1},
	urldate = {2025-11-19},
	journal = {Solar Physics},
	author = {Petrie, Gordon},
	month = jan,
	year = {2017},
	pages = {13},
}

@article{petrie_solar_2022,
	title = {Solar Polar Magnetic Fields: Comparing Full-disk and High-resolution Spectromagnetograph Data},
	volume = {941},
	issn = {0004-637X, 1538-4357},
	shorttitle = {Solar Polar Magnetic Fields},
	url = {https://iopscience.iop.org/article/10.3847/1538-4357/aca1a8},
	doi = {10.3847/1538-4357/aca1a8},
	language = {en},
	number = {2},
	urldate = {2025-11-19},
	journal = {ApJ},
	author = {Petrie, Gordon J. D.},
	month = dec,
	year = {2022},
	pages = {142},
}

@article{yang_long-term_2024,
	title = {Long-term {Variation} of the {Solar} {Polar} {Magnetic} {Fields} at {Different} {Latitudes}},
	volume = {24},
	issn = {2397-6209},
	url = {https://iopscience.iop.org/article/10.1088/1674-4527/ad539a},
	doi = {10.1088/1674-4527/ad539a},
	language = {en},
	number = {7},
	urldate = {2025-11-26},
	journal = {Res. Astron. Astrophys.},
	author = {Yang, Shuhong and Jiang, Jie and Wang, Zifan and Hou, Yijun and Jin, Chunlan and Song, Qiao and Luo, Yukun and Li, Ting and Zhang, Jun and Zhang, Yuzong and Zhou, Guiping and Deng, Yuanyong and Wang, Jingxiu},
	month = jul,
	year = {2024},
	pages = {075015},
}

@article{janardhan_solar_2018,
	title = {Solar cycle 24: {An} unusual polar field reversal},
	volume = {618},
	copyright = {https://www.edpsciences.org/en/authors/copyright-and-licensing},
	issn = {0004-6361, 1432-0746},
	shorttitle = {Solar cycle 24},
	url = {https://www.aanda.org/10.1051/0004-6361/201832981},
	doi = {10.1051/0004-6361/201832981},
	language = {en},
	urldate = {2025-11-19},
	journal = {Astronomy \& Astrophysics},
	author = {Janardhan, P. and Fujiki, K. and Ingale, M. and Bisoi, S. K. and Rout, D.},
	month = oct,
	year = {2018},
	pages = {A148},
}

@article{yang_meridional_2024,
	title = {Meridional Flow in the Solar Polar Caps Revealed by Magnetic Field Observation and Simulation},
	volume = {970},
	issn = {1538-4357},
	url = {https://iopscience.iop.org/article/10.3847/1538-4357/ad61e2},
	doi = {10.3847/1538-4357/ad61e2},
	language = {en},
	number = {2},
	urldate = {2025-11-26},
	journal = {ApJ},
	author = {Yang, Shuhong and Jiang, Jie and Wang, Zifan and Hou, Yijun and Jin, Chunlan and Song, Qiao and Luo, Yukun and Li, Ting and Zhang, Jun and Zhang, Yuzong and Zhou, Guiping and Deng, Yuanyong and Wang, Jingxiu},
	month = aug,
	year = {2024},
	pages = {183},
}

@article{tsuneta_solar_2008,
	title = {The {Solar} {Optical} {Telescope} for the {Hinode} {Mission}: {An} {Overview}},
	volume = {249},
	copyright = {http://www.springer.com/tdm},
	issn = {0038-0938, 1573-093X},
	shorttitle = {The {Solar} {Optical} {Telescope} for the {Hinode} {Mission}},
	url = {http://link.springer.com/10.1007/s11207-008-9174-z},
	doi = {10.1007/s11207-008-9174-z},
	language = {en},
	number = {2},
	urldate = {2025-11-19},
	journal = {Sol Phys},
	author = {Tsuneta, S. and Ichimoto, K. and Katsukawa, Y. and Nagata, S. and Otsubo, M. and Shimizu, T. and Suematsu, Y. and Nakagiri, M. and Noguchi, M. and Tarbell, T. and Title, A. and Shine, R. and Rosenberg, W. and Hoffmann, C. and Jurcevich, B. and Kushner, G. and Levay, M. and Lites, B. and Elmore, D. and Matsushita, T. and Kawaguchi, N. and Saito, H. and Mikami, I. and Hill, L. D. and Owens, J. K.},
	month = jun,
	year = {2008},
	pages = {167--196},
}

@article{tsuneta_magnetic_2008,
	title = {The {Magnetic} {Landscape} of the {Sun}'s {Polar} {Region}},
	volume = {688},
	issn = {0004-637X, 1538-4357},
	url = {https://iopscience.iop.org/article/10.1086/592226},
	doi = {10.1086/592226},
	language = {en},
	number = {2},
	urldate = {2025-11-19},
	journal = {Astrophysical Journal},
	author = {Tsuneta, S. and Ichimoto, K. and Katsukawa, Y. and Lites, B. W. and Matsuzaki, K. and Nagata, S. and Orozco Suárez, D. and Shimizu, T. and Shimojo, M. and Shine, R. A. and Suematsu, Y. and Suzuki, T. K. and Tarbell, T. D. and Title, A. M.},
	month = dec,
	year = {2008},
	pages = {1374--1381}
}

@article{ito_is_2010,
	title = {{IS} {THE} {POLAR} {REGION} {DIFFERENT} {FROM} {THE} {QUIET} {REGION} {OF} {THE} {SUN}?},
	volume = {719},
	issn = {0004-637X, 1538-4357},
	url = {https://iopscience.iop.org/article/10.1088/0004-637X/719/1/131},
	doi = {10.1088/0004-637X/719/1/131},
	language = {en},
	number = {1},
	urldate = {2025-11-25},
	journal = {ApJ},
	author = {Ito, Hiroaki and Tsuneta, Saku and Shiota, Daikou and Tokumaru, Munetoshi and Fujiki, Ken'ichi},
	month = aug,
	year = {2010},
	pages = {131--142},
}

@article{orozco_suarez_quiet-sun_2007,
	title = {Quiet-{Sun} {Internetwork} {Magnetic} {Fields} from the {Inversion} of \textit{{Hinode}} {Measurements}},
	volume = {670},
	issn = {0004-637X, 1538-4357},
	url = {https://iopscience.iop.org/article/10.1086/524139},
	doi = {10.1086/524139},
	language = {en},
	number = {1},
	urldate = {2025-11-19},
	journal = {The Astrophysical Journal},
	author = {Orozco Suárez, D. and Bellot Rubio, L. R. and Del Toro Iniesta, J. C. and Tsuneta, S. and Lites, B. W. and Ichimoto, K. and Katsukawa, Y. and Nagata, S. and Shimizu, T. and Shine, R. A. and Suematsu, Y. and Tarbell, T. D. and Title, A. M.},
	month = nov,
	year = {2007},
	pages = {L61--L64},
}

@article{kubo_comparison_2025,
	title = {Comparison of {Polar} {Magnetic} {Fields} {Derived} from {MILOS} and {MERLIN} {Inversions} with {Hinode}/{SOT}-{SP} {Data}},
	volume = {300},
	issn = {0038-0938, 1573-093X},
	url = {https://link.springer.com/10.1007/s11207-025-02487-z},
	doi = {10.1007/s11207-025-02487-z},
	language = {en},
	number = {5},
	urldate = {2025-11-25},
	journal = {Sol Phys},
	author = {Kubo, Masahito and Shiota, Daikou and Katsukawa, Yukio and Shimojo, Masumi and Orozco Suárez, David and Nitta, Nariaki and DeRosa, Marc and Centeno, Rebecca and Iijima, Haruhisa and Matsumoto, Takuma and Masuda, Satoshi},
	month = may,
	year = {2025},
	pages = {69},
}

@article{petrovay_solar_2020,
	title = {Solar cycle prediction},
	volume = {17},
	issn = {2367-3648, 1614-4961},
	url = {http://link.springer.com/10.1007/s41116-020-0022-z},
	doi = {10.1007/s41116-020-0022-z},
	language = {en},
	number = {1},
	urldate = {2025-11-19},
	journal = {Living Rev Sol Phys},
	author = {Petrovay, Kristóf},
	month = dec,
	year = {2020},
	pages = {2},
}

@article{petrie_solar_2015,
	title = {Solar {Magnetism} in the {Polar} {Regions}},
	volume = {12},
	issn = {2367-3648, 1614-4961},
	url = {http://link.springer.com/10.1007/lrsp-2015-5},
	doi = {10.1007/lrsp-2015-5},
	language = {en},
	number = {1},
	urldate = {2025-11-19},
	journal = {Living Rev. Sol. Phys.},
	author = {Petrie, Gordon J. D.},
	month = dec,
	year = {2015},
	pages = {5},
}

@article{yang_variations_2025,
	title = {Variations of the {Vector} {Magnetic} {Structures} in the {Solar} {Polar} {Regions} {Observed} by {Hinode}},
	volume = {993},
	issn = {0004-637X, 1538-4357},
	url = {https://iopscience.iop.org/article/10.3847/1538-4357/ae0b61},
	doi = {10.3847/1538-4357/ae0b61},
	language = {en},
	number = {1},
	urldate = {2025-11-27},
	journal = {ApJ},
	author = {Yang, Shuhong and Jin, Chunlan and Song, Qiao and Zhang, Yuzong and Li, Yin and Hou, Yijun and Li, Ting and Zhou, Guiping and Deng, Yuanyong and Wang, Jingxiu},
	month = nov,
	year = {2025},
	pages = {9},
}

@article{pesnell_solar_2012,
	title = {The {Solar} {Dynamics} {Observatory} ({SDO})},
	volume = {275},
	issn = {0038-0938, 1573-093X},
	url = {http://link.springer.com/10.1007/s11207-011-9841-3},
	doi = {10.1007/s11207-011-9841-3},
	language = {en},
	number = {1-2},
	urldate = {2025-11-28},
	journal = {Sol Phys},
	author = {Pesnell, W. Dean and Thompson, B. J. and Chamberlin, P. C.},
	month = jan,
	year = {2012},
	pages = {3--15},
}

@article{lites_b_suite_2007,
	title = {A suite of community tools for spectro-polarimetric analysis},
	volume = {78},
	journal = {MmSAI},
	author = {{Lites, B.} and {Casini, R.} and {Garcia, J.} and {Socas-Navarro, H.}},
	year = {2007},
	pages = {148},
}

@article{deng_vector_1999,
	title = {Vector magnetic field in solar polar region},
	volume = {42},
	copyright = {http://www.springer.com/tdm},
	issn = {1006-9283, 1862-2763},
	url = {http://link.springer.com/10.1007/BF02889512},
	doi = {10.1007/BF02889512},
	language = {en},
	number = {10},
	urldate = {2025-11-19},
	journal = {Science in China Series A: Mathematics},
	author = {Deng, Yuanyong and Wang, Jingxiu and Ai, Guoxiang},
	month = oct,
	year = {1999},
	pages = {1096--1102},
}

@article{howard_solar_1990,
	title = {Solar surface velocity fields determined from small magnetic features},
	volume = {130},
	issn = {1573-093X},
	url = {https://doi.org/10.1007/BF00156795},
	doi = {10.1007/BF00156795},
	number = {1},
	journal = {Solar Physics},
	author = {Howard, R. F. and Harvey, J. W. and Forgach, S.},
	month = dec,
	year = {1990},
	pages = {295--311},
}

@article{jiang_magnetic_2014,
	title = {Magnetic {Flux} {Transport} at the {Solar} {Surface}},
	volume = {186},
	issn = {0038-6308, 1572-9672},
	url = {http://link.springer.com/10.1007/s11214-014-0083-1},
	doi = {10.1007/s11214-014-0083-1},
	language = {en},
	number = {1-4},
	urldate = {2025-12-12},
	journal = {Space Sci Rev},
	author = {Jiang, J. and Hathaway, D. H. and Cameron, R. H. and Solanki, S. K. and Gizon, L. and Upton, L.},
	month = dec,
	year = {2014},
	pages = {491--523},
}

@article{leighton_transport_1964,
	title = {Transport of {Magnetic} {Fields} on the {Sun}.},
	volume = {140},
	issn = {0004-637X, 1538-4357},
	url = {http://adsabs.harvard.edu/doi/10.1086/148058},
	doi = {10.1086/148058},
	language = {en},
	urldate = {2025-12-12},
	journal = {ApJ},
	author = {Leighton, Robert B.},
	month = nov,
	year = {1964},
	pages = {1547},
}

@article{bilenko_solar_2026,
	title = {Solar {Polar} {Field} {Reversals} as the {Result} of the {Global} {Magnetic} {Field} {Meridional} {Flows}},
	volume = {301},
	issn = {0038-0938, 1573-093X},
	url = {https://link.springer.com/10.1007/s11207-026-02627-z},
	doi = {10.1007/s11207-026-02627-z},
	language = {en},
	number = {3},
	urldate = {2026-04-23},
	journal = {Solar Physics},
	author = {Bilenko, Irina A.},
	month = mar,
	year = {2026},
	pages = {42},
}

@article{lin_high-resolution_1994,
	title = {High-resolution observations of the polar magnetic fields of the sun},
	volume = {155},
	copyright = {http://www.springer.com/tdm},
	issn = {0038-0938, 1573-093X},
	url = {http://link.springer.com/10.1007/BF00680594},
	doi = {10.1007/BF00680594},
	language = {en},
	number = {2},
	urldate = {2025-11-19},
	journal = {Sol Phys},
	author = {Lin, H. and Varsik, J. and Zirin, H.},
	month = dec,
	year = {1994},
	pages = {243--256},
}

@article{pishkalo_polar_2019,
	title = {On {Polar} {Magnetic} {Field} {Reversal} in {Solar} {Cycles} 21, 22, 23, and 24},
	volume = {294},
	issn = {0038-0938, 1573-093X},
	url = {http://link.springer.com/10.1007/s11207-019-1520-9},
	doi = {10.1007/s11207-019-1520-9},
	language = {en},
	number = {10},
	urldate = {2025-11-19},
	journal = {Sol Phys},
	author = {Pishkalo, Mykola I.},
	month = oct,
	year = {2019},
	pages = {137},
}

@article{sun_polar_2015,
	title = {{ON} {POLAR} {MAGNETIC} {FIELD} {REVERSAL} {AND} {SURFACE} {FLUX} {TRANSPORT} {DURING} {SOLAR} {CYCLE} 24},
	volume = {798},
	issn = {1538-4357},
	url = {https://iopscience.iop.org/article/10.1088/0004-637X/798/2/114},
	doi = {10.1088/0004-637X/798/2/114},
	language = {en},
	number = {2},
	urldate = {2025-11-19},
	journal = {ApJ},
	author = {Sun, Xudong and Todd Hoeksema, J. and Liu, Yang and Zhao, Junwei},
	month = jan,
	year = {2015},
	pages = {114},
}

@article{petrie_polar_2023,
	title = {Polar {Photospheric} {Magnetic} {Field} {Evolution} and {Global} {Flux} {Transport}},
	volume = {298},
	issn = {1573-093X},
	url = {https://link.springer.com/10.1007/s11207-023-02134-5},
	doi = {10.1007/s11207-023-02134-5},
	language = {en},
	number = {3},
	urldate = {2025-11-19},
	journal = {Sol Phys},
	author = {Petrie, Gordon J. D. },
	month = mar,
	year = {2023},
	pages = {43},
}

@article{clette_recalibration_2023,
	title = {Recalibration of the {Sunspot}-{Number}: {Status} {Report}},
	volume = {298},
	issn = {0038-0938, 1573-093X},
	shorttitle = {Recalibration of the {Sunspot}-{Number}},
	url = {https://link.springer.com/10.1007/s11207-023-02136-3},
	doi = {10.1007/s11207-023-02136-3},
	language = {en},
	number = {3},
	urldate = {2025-12-15},
	journal = {Sol Phys},
	author = {Clette, F. and Lefèvre, L. and Chatzistergos, T. and Hayakawa, H. and Carrasco, V. M. S. and Arlt, R. and Cliver, E. W. and Dudok de Wit, T. and Friedli, T. K. and Karachik, N. and Kopp, G. and Lockwood, M. and Mathieu, S. and Muñoz-Jaramillo, A. and Owens, M. and Pesnell, D. and Pevtsov, A. and Svalgaard, L. and Usoskin, I. G. and Van Driel-Gesztelyi, L. and Vaquero, J. M.},
	month = mar,
	year = {2023},
	pages = {44},
}

@article{cameron_solar_2016,
	title = {{SOLAR} {CYCLE} 25: {ANOTHER} {MODERATE} {CYCLE}?},
	volume = {823},
	issn = {2041-8205, 2041-8213},
	shorttitle = {{SOLAR} {CYCLE} 25},
	url = {https://iopscience.iop.org/article/10.3847/2041-8205/823/2/L22},
	doi = {10.3847/2041-8205/823/2/L22},
	language = {en},
	number = {2},
	urldate = {2025-12-11},
	journal = {ApJL},
	author = {Cameron, R. H. and Jiang, J. and Schüssler, M.},
	month = jun,
	year = {2016},
	pages = {L22},
}

@article{kumar_polar_2021,
	title = {The {Polar} {Precursor} {Method} for {Solar} {Cycle} {Prediction}: {Comparison} of {Predictors} and {Their} {Temporal} {Range}},
	volume = {909},
	issn = {0004-637X, 1538-4357},
	shorttitle = {The {Polar} {Precursor} {Method} for {Solar} {Cycle} {Prediction}},
	url = {https://iopscience.iop.org/article/10.3847/1538-4357/abdbb4},
	doi = {10.3847/1538-4357/abdbb4},
	language = {en},
	number = {1},
	urldate = {2025-12-16},
	journal = {ApJ},
	author = {Kumar, Pawan and Nagy, Melinda and Lemerle, Alexandre and Binay Karak, Bidya and Petrovay, Kristof},
	month = mar,
	year = {2021},
	pages = {87},
}

@article{lites_hinode_2013,
	title = {The {Hinode} {Spectro}-{Polarimeter}},
	volume = {283},
	copyright = {http://creativecommons.org/licenses/by/2.0},
	issn = {0038-0938, 1573-093X},
	url = {http://link.springer.com/10.1007/s11207-012-0206-3},
	doi = {10.1007/s11207-012-0206-3},
	language = {en},
	number = {2},
	urldate = {2025-12-17},
	journal = {Solar Physics},
	author = {Lites, B. W. and Akin, D. L. and Card, G. and Cruz, T. and Duncan, D. W. and Edwards, C. G. and Elmore, D. F. and Hoffmann, C. and Katsukawa, Y. and Katz, N. and Kubo, M. and Ichimoto, K. and Shimizu, T. and Shine, R. A. and Streander, K. V. and Suematsu, A. and Tarbell, T. D. and Title, A. M. and Tsuneta, S.},
	month = apr,
	year = {2013},
	pages = {579--599},
}

@article{lites_sp_prep_2013,
	title = {The {SP}\_PREP {Data} {Preparation} {Package} for the {Hinode} {Spectro}-{Polarimeter}},
	volume = {283},
	copyright = {http://www.springer.com/tdm},
	issn = {0038-0938, 1573-093X},
	url = {http://link.springer.com/10.1007/s11207-012-0205-4},
	doi = {10.1007/s11207-012-0205-4},
	language = {en},
	number = {2},
	urldate = {2025-12-17},
	journal = {Sol Phys},
	author = {Lites, B. W. and Ichimoto, K.},
	month = apr,
	year = {2013},
	pages = {601--629},
}

@article{scherrer_helioseismic_2012,
	title = {The {Helioseismic} and {Magnetic} {Imager} ({HMI}) {Investigation} for the {Solar} {Dynamics} {Observatory} ({SDO})},
	volume = {275},
	copyright = {https://creativecommons.org/licenses/by-nc/2.0},
	issn = {0038-0938, 1573-093X},
	url = {http://link.springer.com/10.1007/s11207-011-9834-2},
	doi = {10.1007/s11207-011-9834-2},
	language = {en},
	number = {1-2},
	urldate = {2026-01-09},
	journal = {Solar Physics},
	author = {Scherrer, P. H. and Schou, J. and Bush, R. I. and Kosovichev, A. G. and Bogart, R. S. and Hoeksema, J. T. and Liu, Y. and Duvall, T. L. and Zhao, J. and Title, A. M. and Schrijver, C. J. and Tarbell, T. D. and Tomczyk, S.},
	month = jan,
	year = {2012},
	pages = {207--227},
}

@article{schou_design_2012,
	title = {Design and {Ground} {Calibration} of the {Helioseismic} and {Magnetic} {Imager} ({HMI}) {Instrument} on the {Solar} {Dynamics} {Observatory} ({SDO})},
	volume = {275},
	copyright = {https://creativecommons.org/licenses/by-nc/2.0},
	issn = {0038-0938, 1573-093X},
	url = {http://link.springer.com/10.1007/s11207-011-9842-2},
	doi = {10.1007/s11207-011-9842-2},
	language = {en},
	number = {1-2},
	urldate = {2026-01-09},
	journal = {Solar Physics},
	author = {Schou, J. and Scherrer, P. H. and Bush, R. I. and Wachter, R. and Couvidat, S. and Rabello-Soares, M. C. and Bogart, R. S. and Hoeksema, J. T. and Liu, Y. and Duvall, T. L. and Akin, D. J. and Allard, B. A. and Miles, J. W. and Rairden, R. and Shine, R. A. and Tarbell, T. D. and Title, A. M. and Wolfson, C. J. and Elmore, D. F. and Norton, A. A. and Tomczyk, S.},
	month = jan,
	year = {2012},
	pages = {229--259},
}

@article{van_ballegooijen_magnetic_1998,
	title = {Magnetic {Flux} {Transport} and the {Formation} of {Filament} {Channels} on the {Sun}},
	volume = {501},
	issn = {0004-637X, 1538-4357},
	url = {https://iopscience.iop.org/article/10.1086/305823},
	doi = {10.1086/305823},
	language = {en},
	number = {2},
	urldate = {2026-01-24},
	journal = {The Astrophysical Journal},
	author = {Van Ballegooijen, A. A. and Cartledge, N. P. and Priest, E. R.},
	month = jul,
	year = {1998},
	pages = {866--881},
}

@article{wang_surface_2017,
	title = {Surface {Flux} {Transport} and the {Evolution} of the {Sun}’s {Polar} {Fields}},
	volume = {210},
	issn = {0038-6308, 1572-9672},
	url = {http://link.springer.com/10.1007/s11214-016-0257-0},
	doi = {10.1007/s11214-016-0257-0},
	language = {en},
	number = {1-4},
	urldate = {2026-03-10},
	journal = {Space Science Reviews},
	author = {Wang, Y.-M.},
	month = sep,
	year = {2017},
	pages = {351--365},
}

@article{calchetti_first_2025,
	title = {The {First} out-of-{Ecliptic} {Observations} of the {Polar} {Magnetic} {Field} of the {Sun}},
	volume = {995},
	issn = {2041-8205, 2041-8213},
	url = {https://iopscience.iop.org/article/10.3847/2041-8213/ae2850},
	doi = {10.3847/2041-8213/ae2850},
	language = {en},
	number = {2},
	urldate = {2025-12-22},
	journal = {The Astrophysical Journal Letters},
	author = {Calchetti, D. and Solanki, S. K. and Hirzberger, J. and Valori, G. and Chitta, L. P. and Blanco Rodríguez, J. and Giunta, A. and Grundy, T. and Albert, K. and Appourchaux, T. and Bailén, F. J. and Bellot Rubio, L. R. and Feller, A. and Gandorfer, A. and Gizon, L. and Korpi-Lagg, A. and Li, X. and Moreno Vacas, A. and Oba, T. and Orozco Suárez, D. and Schou, J. and Schühle, U. and Sinjan, J. and Strecker, H. and Del Toro Iniesta, J. C. and Ulyanov, A. and Volkmer, R. and Woch, J.},
	month = dec,
	year = {2025},
	pages = {L60},
}

@article{liu_comparison_2012,
	title = {Comparison of {Line}-of-{Sight} {Magnetograms} {Taken} by the {Solar} {Dynamics} {Observatory}/{Helioseismic} and {Magnetic} {Imager} and {Solar} and {Heliospheric} {Observatory}/{Michelson} {Doppler} {Imager}},
	volume = {279},
	copyright = {http://www.springer.com/tdm},
	issn = {0038-0938, 1573-093X},
	url = {http://link.springer.com/10.1007/s11207-012-9976-x},
	doi = {10.1007/s11207-012-9976-x},
	language = {en},
	number = {1},
	urldate = {2026-03-31},
	journal = {Solar Physics},
	author = {Liu, Y. and Hoeksema, J. T. and Scherrer, P. H. and Schou, J. and Couvidat, S. and Bush, R. I. and Duvall, T. L. and Hayashi, K. and Sun, X. and Zhao, X.},
	month = jul,
	year = {2012},
	pages = {295--316},
}

@article{makarov_duration_2003,
	title = {Duration of {Polar} {Activity} {Cycles} and {Their} {Relation} to {Sunspot} {Activity}},
	volume = {214},
	copyright = {https://www.springernature.com/gp/researchers/text-and-data-mining},
	issn = {0038-0938, 1573-093X},
	url = {https://link.springer.com/10.1023/A:1024003708284},
	doi = {10.1023/A:1024003708284},
	language = {en},
	number = {1},
	urldate = {2026-06-04},
	journal = {Solar Physics},
	author = {Makarov, V.I. and Tlatov, A.G. and Sivaraman, K.R.},
	month = may,
	year = {2003},
	pages = {41--54},
}

@ARTICLE{1986BASI,
       author = {{Makarov}, V.~I. and {Sivaraman}, K.~R.},
        title = "{On the epochs of polarity reversals of the polar magnetic field of the sun during 1870-1982}",
      journal = {Bulletin of the Astronomical Society of India},
         year = 1986,
        month = sep,
       volume = {14},
        pages = {163-167},
       adsurl = {https://ui.adsabs.harvard.edu/abs/1986BASI...14..163M}
}

\end{document}